\documentclass[a4paper,11pt]{article}
\usepackage[usenames]{color}

\usepackage{amsfonts,amssymb}
\usepackage{theorem}
\usepackage{cite}
\usepackage{epsfig}

\def\G{\Gamma}

\newcommand{\comment}[1]{}

\begin{document}
\Large
{\bf
\centerline{On the Weak Coupling Limit for Massive Yang-Mills}}

\large
\rm
\vskip 0.5 truecm

\begin{center}
Daniele~Bettinelli$^{a,}$\footnote{e-mail: {\tt daniele.bettinelli@mi.infn.it}}
and 
Ruggero~Ferrari$^{a,b,}$\footnote{e-mail: {\tt ruggero.ferrari@mi.infn.it}}
\end{center}

\small
\medskip
\begin{center}
$^a$
Dip. di Fisica, Universit\`a degli Studi di Milano\\
and INFN, Sezione di Milano\\
via Celoria 16, I-20133 Milano, Italy\\
and\\
$^b$
Center for Theoretical Physics\\
Laboratory for Nuclear Science\\
and Department of Physics\\
Massachusetts Institute of Technology\\
Cambridge, Massachusetts 02139\\
(MIT-CTP 4372, IFUM-992-FT, September, 2012 )
\end{center}

\normalsize

\vskip 1.0 truecm
\normalsize

\begin{quotation}
\rm{\Large Abstract:}
For small values of the gauge 
coupling constant, we compare the densities of the energy of the vacuum 
and of the order parameter, 
evaluated in the lattice Monte Carlo simulation
and  in the perturbative field theory at two loop (Minkowski). 
The continuum calculation allows a very good fit of the
simulation results, away from the phase transition line.  
This confirms the conjecture that the lattice
provides a regularization of the (nonrenormalizable) massive Yang-Mills
and moreover it shows the physical meaning of the parameters used in the
simulation.
\end{quotation}
\newpage
%%%%%%%%%%%%%%%%%%%%%%%%%%%%%%%%%%%%%%%%%%%%%%%%%%%%
%%%%%%%%%%%%%%%%%%%%%%%%%%%%%%%%%%%%%%%%%%%%%%%%%%%%
\normalsize
\rm

%%%%%%%%%%%%%%%%%%%%%%%%%%%%%%%%%%%%%%%%%%%%%
\section{Introduction}
\label{sec:intro}
%%%%%%%%%%%%%%%%%%%%%%%%%%%%%%%%%%%%%%%%%%%%%%
%
In a recent paper \cite{Ferrari:2011aa}, a lattice gauge theory for
massive $SU(2)$ Yang-Mills  in dimension four has been proposed. 
The completely gauge
independent analysis shows the presence of a Transition Line (TL)
in the parameter space $(m^2\sim 1/ \beta)$, which seems to 
indicate a phase transition (but it might also be a cross over)  
for large $\beta$, starting from
$\beta\sim 2.2$. The large $\beta$ limit value of $\beta m_c^2$,
where $m_c^2$ is the critical value of $m^2$ on the TL,
is in good agreement with the
critical point of the nonlinear sigma model in four dimensions quoted in Ref.\cite{Baaquie:1992cw}, 
namely $ 0.65\pm 0.04$.
\par
The above-mentioned lattice gauge model has been studied previously (see 
\cite{Fradkin:1978dv}-\cite{Bonati:2009pf}) as an example
of Higgs mechanism with a frozen length. In Ref. \cite{Ferrari:2011aa} 
we showed that it is the perfect
tool for the simulation of the massive Yang-Mills (i.e. with a mass 
{\sl \`a la } St\"uckelberg). The phase TL has an end-point around 
( $\beta\sim 2.2, \,\, m^2\sim 0.381$): for smaller
$\beta$ there is a smooth transition from one phase to the other, 
while for larger $\beta$ there are numerical indications of
singularities in the derivatives  with
respect to $m^2$ and to $\beta$  of the energy
and of the order parameter 
(i.e. the $\ln m^2$ derivative of the free energy).
\par
The lattice model turns out to be extremely interesting by itself
because of the states structure, the dynamical content, the
limit of small gauge coupling, the complexity of the phase
diagram in the parameter space $(\beta, m^2)$. In this paper
we want to compare the lattice model with the Yang-Mills gauge
theory where a mass  {\sl \`a la } St\"uckelberg is introduced.
The latter model is nonrenormalizable, but we have devised a method
to make  a respectable physical theory out of it 
(\cite{Bettinelli:2007tq}-\cite{Bettinelli:2007zn}).  
The comparison of the two models, via analytic
continuation, can provide interesting results, if we find some
physical quantities where they are expected to give identical
results. Global quantities might be good candidates in order to
compare the two models and thus to identify the physical meaning
of the lattice parameters.
\par
To this end, we investigate on the small gauge coupling behavior, 
i.e. large $\beta$,
 of some global observables such as densities of the {\sl ``energy of
the vacuum''} and of the order parameter
\begin{eqnarray}&&
{\cal E}_L = -\frac{1}{N }\beta
\frac{\partial}{ \partial\beta}\ln Z_L
\nonumber\\&&
{\mathfrak  C}_L = -\frac{1}{N } m^2 \frac{\partial}{ \partial m^2}\ln Z_L\, ,
\label{intro.1}
\end{eqnarray}
where $N$ is the number of sites  and 
$Z_L$ the partition function
\footnote{
$\cdot_M$ for Minkowski, $\cdot_E$ for Euclidean and $\cdot_L$ for Lattice.}.
{It should be noticed that in Ref. \cite{Ferrari:2011aa} the order parameter is
defined as
\begin{eqnarray}
1-\frac{\mathfrak{C}_L}{m^2D\beta},
\label{intro.1p.bis}
\end{eqnarray}
in order to put in evidence its behavior for $m^2=0$
and $m^2 \to \infty$. Here these properties are not relevant.}
\par
In this work we discuss two important questions: i) to understand
the phenomenological meaning of the parameters $\beta $ and
$m^2$ (in the naive continuum limit, $a\to 0$ , $\beta= 4 g^{-2}$ and
$m a^{-1}$ is the mass of the vector mesons); ii) to compare
the lattice regularization of the theory with the continuum 
Minkowskian formulation of massive Yang-Mills theory
proposed in Refs. 
\cite{Bettinelli:2007tq} and \cite{Bettinelli:2007cy}.
Regarding i), we want to find the scales necessary in order to interpret 
the otherwise dimensionless parameter $m$. For the point ii),
we assume that the finite lattice artifacts are not 
relevant quantitatively for the global quantities of eq. (\ref{intro.1}). 
\par
The transition from Euclidean
to Minkowskian quantities is performed in the usual conventional
way:  $\exp (iS_M)$ is the weight in the path integral 
and the arrow of time (i.e. anticlockwise Wick rotation)
is chosen in order to match the edge-of-the-wedge theorem 
\cite{Streater:1989vi}.
\par
In the subtraction procedure \cite{Ferrari:2005ii} - \cite{Bettinelli:2007zn}
  a scale $\Lambda$ of the radiative correction is introduced, then it
is natural to look for a correspondence implemented
by the mapping
\begin{eqnarray}&&
g^2 =f(\beta,m^2)
\nonumber\\&&
\frac{M^2}{\Lambda^2} =s(\beta,m^2)
\nonumber\\&&
 Ma = t(\beta,m^2),
\label{intro.2}
\end{eqnarray}
where the parameters $g,M$ are the field theory coupling
constant and mass respectively, while $a$ is the {\sl "lattice spacing''},
 i.e. a length introduced for dimensional reasons.
Eqs. (\ref{intro.2}) show that the mapping of the continuum field
theory onto the lattice implies a choice of a surface in the three-dimensional
space spanned by the dimensionless parameters ($g$, $Ma$ and $\Lambda a$).
The convenience of these choices might be sustained and tested 
on the physical observables to be compared.  
\par
The fit indicates that far from $m_c^2$ ($m^2>m_c^2$) and for large $\beta$
 the field theory predictions at two-loop describe
very well the lattice simulation data and moreover $a\sim M^{-1}$,
$m \sim \Lambda^{-1} M$ and the Yang-Mills coupling constant $g$
is a mildly decreasing function of $\beta$ (the lattice parameter)
with $g^2 \sim 4/\beta$ at $\beta =20$.

%%%%%%%%%%%%%%%%%%%%%%%%%%%%%%%%%%%%%%%%%%%%%%
%%%%%%%%%%%%%%%%%%%%%%%%%%%%%%%%%%%%%%%%%%%%%%
\section{ Massive Yang-Mills in Field Theory (Minkowski) }
\label{sec:YM}
%%%%%%%%%%%%%%%%%%%%%%%%%%%%%%%%%%%%%%%%%%%%%%%%%%
%
In the continuum quantum field theory the classical action is 
\begin{eqnarray}
S_{ M}
   = \frac{\Lambda^{(D-4)}}{g^2} \int d^Dx \, \Big ( - \frac{1}{4} 
                              G_{a\mu\nu}[A] G^{\mu\nu}_a[A] + 
                    \frac{M^2}{2}( A_{a\mu}-F_{a\mu})^2 \Big ) \, .
\label{YM.1}
\end{eqnarray}
The path integral functional $Z_M$  is
obtained by integration over the fields $A_\mu$ and $\Omega \in SU(2)$,
where 
($\tau$ are the Pauli matrices)
\begin{eqnarray}
\Omega = \phi_0 + i \tau_j \phi_j,\qquad F_\mu \equiv
i\Omega \partial_\mu\Omega^\dagger. 
\label{YM.2}
\end{eqnarray}
The constraint on $\Omega$ 
\begin{eqnarray}
1= \phi_0^2 +  \vec\phi^2
\label{YM.3}
\end{eqnarray}
renders the theory in eq. (\ref{YM.1}) nonrenormalizable.
The procedure of subtraction of the infinities in perturbation
theory is described in detail in Ref. \cite{Bettinelli:2007tq}.
Here we account only for the final practical rule: only pure
pole subtraction is performed on the dimensionally regularized 
Feynman amplitudes normalized by
\begin{eqnarray}
\Lambda^{(4-D)}{\cal A}.
\label{YM.4}
\end{eqnarray}
No additional finite adjustment is allowed, e.g. on-shell
normalization.
\par
We use Landau gauge because it is very useful in a massive
theory, where the massless modes are all unphysical.
The absence of the massless modes contributions provides 
a good check for the calculation of the physical observables.
\par
The {\sl ``energy of the vacuum''} per hypercube $a^4$
is then given by the path integral mean value $\langle\rangle_M$
\begin{eqnarray}
{\cal E}_M(g,M,\Lambda,a)=\Big\langle - \frac{a^4}{g^2}  \Big ( - \frac{1}{4} 
                              G_{a\mu\nu}[A] G^{\mu\nu}_a[A] +
                    \frac{M^2}{2} A_{a\mu}^2 \dots \Big) \Big\rangle_M,
\label{YM.5}
\end{eqnarray}
where $\dots$ stand for the counterterms and 
for the terms of the classical  action depending
on the unphysical fields as $\Omega$ and Faddeev-Popov ghosts $c,\bar c$.
The full expression is given in Ref. \cite{Bettinelli:2007tq}.
\par
\begin{figure}
\begin{center}
\includegraphics[clip,width=0.8\textwidth,,height=0.1\textheight]
{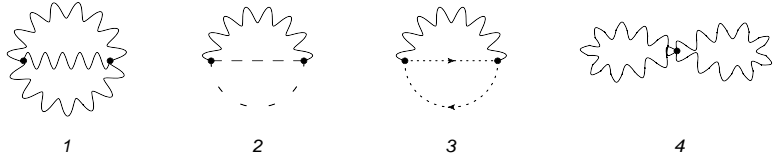}
\end{center}
\caption{The Goldstone boson lines are dashed. The Faddeev-Popov
propagators are dotted.}
\label{fig.3}
\end{figure}

The ${\cal O}(g^0)$ of ${\cal E}_M$ is zero. This result can be proved
formally by applying the operator $g \frac{\partial}{\partial g}$
to $Z_M$ before and after the rescaling of the field $A_\mu \to g A_\mu$.
The ${\cal O}(g^2)$ part amounts to the evaluation of the graphs
in Fig.  \ref{fig.3}; i.e. after rescaling
\begin{eqnarray}
{\cal E}_M =\Big\langle  a^4 \frac{g}{2} \frac{\partial}{\partial g}  
\Big ( - \frac{1}{4} 
                              G_{a\mu\nu}[A] G^{\mu\nu}_a[A] +
                    \frac{M^2}{2} A_{a\mu}^2 \dots \Big) \Big\rangle_M.
\label{YM.6}
\end{eqnarray}
%

%%%%%%%%%%%%%%%%%%%%%%%%%%%%%%%%%%%%%
\subsection{Two Loop Amplitude}
%%%%%%%%%%%%%%%%%%%%%%%%%%%%%%%%%%%%%
%
A straightforward calculation gives the Feynman amplitude 
properly normalized \cite{Ferrari:2005ii} by the factor $\Lambda^{(4-D)}$
\begin{eqnarray}
{\cal E}_{1M}=
-\frac{3}{2}a^4 g^2 \Bigg[3\Big(D-\frac{5}{4}\Big) M^2 I_2[M]-\Big(D^2-4D+\frac{15}{4}\Big) A_0[M]^2 \Bigg].
\label{intr.1}
\end{eqnarray}
The integrals are defined by
\begin{eqnarray}&&
 I_2[M]= \int_M \frac{d^Dq}{(2\pi)^D}\int_M \frac{d^Dp}{(2\pi)^D}
 \frac{\Lambda^{(8-2D)}}{(q^2-M^2)(p^2-M^2)[(q+p)^2-M^2]}
\nonumber\\&&
A_0[M]=
\int_M \frac{d^Dq}{(2\pi)^D}\frac{\Lambda^{(4-D)}}{(q^2-M^2)}.
\label{intr.1.1}
\end{eqnarray}
It should be noticed that all contributions by the massless modes
cancel out (no contributions from the unphysical sector).
\par 
The tadpole integral is
\begin{eqnarray}
&&\!\!\!\!\!\!\!\!\!\!\!\!\!\!\!\!\!\!\!
A_0[M] = i\frac{M^2}{(4\pi)^2}\Bigg[-\frac{2}{D-4} -\ln \Delta  
\nonumber\\
&&~~~~~~~~~~~~~
-\frac{1}{4}(D-4) \Big(\ln^2 \Delta + 1  + \frac{\pi^2}{6} \Big)\Bigg] + {\cal O}\big((D-4)^2\big),
\label{form.6}
\end{eqnarray}
where
\begin{eqnarray}
\Delta = \frac{ e^{(\gamma -1)} M^2}{4\pi \Lambda^2}.
\label{form.5}
\end{eqnarray}
Consequently, one gets
\begin{eqnarray}
%&&
A_0[M]^2
= - \frac{M^4}{(4\pi)^4} \Bigg[\frac{4}{(D-4)^2} + \frac{4}{D-4} \ln \Delta + 1 
+ \frac{\pi^2}{6} + 2 \ln^2 \Delta \Bigg].
\label{ext.2}
\end{eqnarray}
The two-loop integral can be found in the literature 
(see Ref. \cite{Davydychev:1995mq})
\begin{eqnarray}
&&\!\!\!\!\!\!\!\!\!\!\!\!\!\!\!\!\!\!\!\!\!
I_2[M] = 3 \frac{M^2}{(4\pi)^4} \Bigg[-\frac{2}{(D-4)^{2}}
+ \frac{1}{D-4}\Big(1-2 \ln \Delta \Big) 
\nonumber\\
&&~~~~~~~~~~
- \Big(  -\ln \Delta  
+ \frac{\pi^2}{12}+\frac{3}{2}+\ln^2 \Delta \Big)
+ \frac{2}{\sqrt 3} Cl_2\Big(\frac{\pi}{3}\Big) \Bigg],
\label{contr.6}
\end{eqnarray}
where the Euler-Mascheroni constant and the Clausen function are
\begin{eqnarray}&&
\gamma = 0.5772156649
\nonumber\\&&
Cl_2\Big(\frac{\pi}{3}\Big)= 1.014941606.
\label{form.7pp}
\end{eqnarray}
Finally, the contributions of the graphs in Fig. \ref{fig.3} add up to
\begin{eqnarray}
&&\!\!\!\!\!\!\!\!\!\!\!\!\!
{\cal E}_{1M}
= -\frac{3}{2} a^4 g^2 \frac{M^4}{(4\pi)^4}\Bigg[
-\frac{69}{2}\frac{1}{(D-4)^2} -\frac{1}{(D-4)} \Big(-\frac{91}{4} + \frac{69}{2}\ln \Delta\Big)
\nonumber\\
&&~~~~
-\frac{163}{8} + \frac{91}{4} \ln \Delta - \frac{69}{4} \ln^2 \Delta 
-\frac{23}{16} \pi^2 + \frac{99}{2\sqrt 3}Cl_2\Big(\frac{\pi}{3}\Big)\Bigg].
\label{sum.1}
\end{eqnarray}
%
%%%%%%%%%%%%%%%%%%%%%%%%%%%%%%%%%%%%%%%%%%%
\subsection{Contribution of the Counterterms}
%%%%%%%%%%%%%%%%%%%%%%%%%%%%%%%%%%%%%%%%%%%%
%
In order to evaluate the contribution of the counterterms,
 we use the results of Ref. \cite{Bettinelli:2007cy} eq. (29)
\begin{eqnarray}
&&\!\!\!\!\!\!\!\!\!\!\!\!
\widehat \G^{(1)} \equiv \int d^D x \, \, \widehat\gamma^{(1)} (x)
\nonumber \\&&  
= \frac{\Lambda^{(D-4)}}{(4\pi)^2} \frac{1}{D-4} \int d^D x
\Big [  \frac{17}{4} \Big(\partial_\mu A_{a\nu} \partial^\mu A_{a}^\nu- (\partial A_a)^2\Big) 
+ \frac{3}{2}M^2 A_a^2   \Big ].
\label{oc.3}
\end{eqnarray}
Then the contribution of the counterterms is
\begin{eqnarray}
&&\!\!\!\!\!\!\!\!\!\!\!\!
{\cal E}_{2M}=
-i \frac{a^4 g^2 M^2}{(4\pi)^2} \frac{D-1}{D-4} \frac{69}{4} A_0[M^2]
\nonumber\\&&
= - \frac{3}{2} 
\frac{a^4 g^2 M^4}{(4\pi)^4} 
\Bigg[\frac{69}{(D-4)^2} + \frac{1}{D-4} \Big(23+\frac{69}{2}\ln \Delta \Big) +
\frac{23}{2} \ln \Delta 
\nonumber\\&&   
~~~~~~~~~~~~~~~~~~~
+ \frac{69}{8} \Big(\ln^2 \Delta + 1 + \frac{\pi^2}{6} \Big)\Bigg].
\label{oc.4}
\end{eqnarray}
The final result for the energy of the vacuum   to order ${\cal O}(g^2)$ is
\begin{eqnarray}
&&\!\!\!\!\!\!\!\!\!\!\!\!\!\!\!\!\!\!\!\!\!
{\cal E}_M = {\cal E}_{1M}+{\cal E}_{2M}= -\frac{3}{2}a^4 g^2 \frac{M^4}{(4\pi)^4}\Bigg[
\frac{69}{2}\frac{1}{(D-4)^2} +\frac{183}{4}\frac{1}{(D-4)}
\nonumber\\&&~~~~~~~~~~~~~~ 
- \frac{69}{8}\ln^2 \Delta 
+ \frac{137}{4}\ln \Delta - \frac{47}{4}
+ \frac{99}{2\sqrt 3} Cl_2\Big(\frac{\pi}{3}\Big)\Bigg] .
\label{tot.1}
\end{eqnarray}
Notice that the singular part has no dependence on $\Delta$. The
finite part is then
\begin{eqnarray}
{\cal E}_M = -\frac{3}{2}a^4 g^2 \frac{M^4}{(4\pi)^4}\Bigg[
- \frac{69}{8}\ln^2 \Delta 
+ \frac{137}{4}\ln \Delta - \frac{47}{4}
+ \frac{99}{2\sqrt 3} Cl_2\Big(\frac{\pi}{3}\Big)\Bigg].
\label{tot.2}
\end{eqnarray}
%
%%%%%%%%%%%%%%%%%%%%%%%%%%%%%%%%%%%%
\subsection{The Order Parameter}
%%%%%%%%%%%%%%%%%%%%%%%%%%%%%%%%%%%%
%
The order parameter at ${\cal O}(g^{0})$ is given by
\begin{eqnarray}
{\mathfrak  C}_M^{(0)}= - a^4\Big\langle \frac{M^2}{2} (A_{a\mu}-F_{a\mu})^2 \Big\rangle_M^{(0)}
=  \frac{3}{2}\frac{a^4M^4}{(4\pi)^{2}}\Big(\frac{6}{D-4}+3\ln \Delta+2\Big),
\label{YM.14}
\end{eqnarray}
where the path integral mean value is taken with the free part of the action.
The ${\cal O}(g^{2})$ can be easily evaluated with a trick. We use the fields scaled
in such a way that the coupling constant appears only in the interacting part
of the action. Then we have the identity
\begin{eqnarray}&&
{\mathfrak  C}_M^{(2)}=
\frac{g}{2}\frac{\partial}{\partial g}{\mathfrak  C}_M^{(2)}
= i\frac{M^2}{N}\frac{\partial}{\partial M^2} \frac{g}{2}\frac{\partial}{\partial g}
\ln Z_M
= - M^2 \frac{\partial}{\partial M^2} {\cal E}_M.
\label{YM.15}
\end{eqnarray}
Thus we use the expression in eq. (\ref{tot.1})
\begin{eqnarray}
&&\!\!\!\!\!\!\!\!\!\!\!\!\!\!\!
{\mathfrak  C}_M^{(2)}=
\frac{3}{2} g^2 \frac{a^4M^4}{(4\pi)^4} \Bigg[
 69 \frac{1}{(D-4)^2}  + \frac{183}{2} \frac{1}{D-4}
\nonumber\\&&~~~~~~~~~~~~~~ 
-\frac{69}{4} \ln^2 \Delta +
\frac{205}{4} \ln \Delta
+\frac{43}{4} + 33 \sqrt{3} Cl_2\Big(\frac{\pi}{3}\Big) \Bigg] .
\label{YM.16}
\end{eqnarray}
We notice that once again the singular terms depending on $\Delta$ cancel out
in the final amplitudes.

%%%%%%%%%%%%%%%%%%%%%%%%%%%%%%%%%%%%%%%%%%%%%
\section{Euclidean-Minkowskian Correspondence}
\label{sec:corr}
%%%%%%%%%%%%%%%%%%%%%%%%%%%%%%%%%%%%%%%%%%%%%%
%
The correspondence is established by requiring that
the observables have the same value. 
The mapping is obtained by the substitution
\begin{eqnarray}&&
x_4 = -i x_0
\nonumber\\&&
A_4 = i A_0
\label{corr.1}
\end{eqnarray}
and by performing an anticlockwise Wick rotation on the $x_0$
integration (to match the statement of the edge-of-the-wedge theorem
\cite{Streater:1989vi}).
The generating functionals are obtained by summing over the field
configurations with the weights 
\begin{eqnarray}&&
e^{-S_E}
\nonumber\\&&
e^{iS_M}\textsl{}
\label{corr.2}
\end{eqnarray}
for the Euclidean $\langle \rangle_E$ and Minkowskian $\langle\rangle_M$
mean values, where $S_E$ is the Euclidean action obtained from $S_M$
by using the mapping in eq. (\ref{corr.1}). 
Then we use  the correspondence
\begin{eqnarray}&&
\!\!\!\!\!\!\!\!\!\!\!\!\!\!\!\!\!\!\!\!\!\!\!
\Big\langle \Big(\dots\phi(x_j)\dots A_{\mu_k}(x_k)\dots\Big)\Big\rangle_M
\Big|_{\!\!\!\!\tiny{\begin{array}{l}\\x_{0}=i\,x_{4}\\A_0=-i A_4\end{array}}} 
\!\!\!\!\!\!\!\!\!\!\!\! =
\Big\langle \Big(\dots\phi(x_j)\dots A_{\mu_k}(x_k)\dots\Big)\Big\rangle_E .
\label{corr.9.0}
\end{eqnarray}
\par
In the Minkowski case the field theory is made finite by the procedure
briefly outlined at the beginning of Sec. \ref{sec:YM}. For  the Euclidean
formulation, we want to investigate the possibility to approximate the
amplitudes by a gauge lattice model given by an appropriate action $S_L$
introduced in  Ref. \cite{Ferrari:2011aa} and discussed later on.
In fact, the paper is devoted to this comparison by considering the
global quantities $
{\cal E}_L$ and $
{\mathfrak  C}_L   $ in lattice gauge theory
in the limit of weak coupling ($\beta \gg 1$), where perturbation theory
can be used for the theory in the continuum.

%%%%%%%%%%%%%%%%%%%%%%%%%%%%%%%%%%%%%%%%%%%
%%%%%%%%%%%%%%%%%%%%%%%%%%%%%%%%%%%%%%%%%%%
\section{The Lattice Model}
\label{sec:model}
%%%%%%%%%%%%%%%%%%%%%%%%%%%%%%%%%%%%%%%%%%
%
The action in eq.(\ref{YM.1}) is invariant under $g_{\scriptstyle{\scriptstyle{L}}}(x)\in SU(2)_L$ local-left 
and $g_{\scriptstyle{R}}\in SU(2)_R$ global-right transformations. 
On the lattice, one can implement the same transformations by using
link variables $U(x,\mu)$ and site variables $\Omega(x)\in SU(2)$. 
We have
\begin{eqnarray}
\!
\scriptstyle{SU(2)_L}\left\{
\begin{array}{l}
\Omega'(x) = g_L(x)\Omega(x) \\
U'(x,\mu) = g_L(x) U(x,\mu)  g^\dagger_L(x+\mu)
\end{array} \right. \!\!, ~\,\,
\scriptstyle{SU(2)_R}\left\{
\begin{array}{l}\Omega'(x) = \Omega(x)g_R^\dagger
\\  U'(x,\mu) = U(x,\mu)
\end{array} \right. \!.
\label{stck.5}
\end{eqnarray}
The lattice model is constructed by assuming the same invariance properties.
The nearest neighbor interaction 
is required to map   na{\"\i}vely  into the action (\ref{YM.1}) 
for zero lattice spacing. The link variable is taken to be
\begin{eqnarray}
U(x,\mu)\simeq \exp(-i a A_\mu(x)).
\label{lat.1}
\end{eqnarray}
Thus the action is
\begin{eqnarray}
S_{ L}=\!\! \frac{\beta}{2} \,\, {\mathfrak Re} \sum_\Box Tr\big\{1-  U_\Box\big\}
+ \frac{\beta}{2} m^2 {\mathfrak Re}
\sum_{x\mu}Tr\Bigl\{1- \Omega(x)^\dagger U(x,\mu)\Omega(x+\mu)\Bigr\},
\label{lat.2}
\end{eqnarray}
where the sum over the plaquette is the Wilson action \cite{Wilson:1974sk}
and the mass term has the (Euclidean) continuum limit
\begin{eqnarray}&&
\frac{\beta}{2} M^2a^2 {\mathfrak Re}
\sum_{x\mu}Tr\Bigl\{1- \Omega(x)^\dagger U(x,\mu)\Omega(x+\mu)\Bigr\}
\nonumber\\&&
\to
\frac{M^2}{g^2} \int d^4x \,\, Tr\,\Big\{(A_\mu - i\Omega\partial_\mu\Omega^\dagger)^2
\Big\},
\label{lat.3}
\end{eqnarray}
to be compared with the corresponding term in eq. (\ref{YM.1}).

%%%%%%%%%%%%%%%%%%%%%%%%%%%%%%%%%%%%%
%%%%%%%%%%%%%%%%%%%%%%%%%%%%%%%%%%%%%
\section{The Lattice Simulation}
\label{sec:sim}
%%%%%%%%%%%%%%%%%%%%%%%%%%%%%%%%%%%%%
%
The partition function is obtained by summing over all configurations given
by the link variables and the $SU(2)$-valued field $\Omega$
\begin{eqnarray}
Z_L[\beta, m^2 , N] = \sum_{\{U,\Omega\}} e^{-S_{ L}},
\label{sim.1}
\end{eqnarray}
where $N$ is the number of sites. \par
In principle the integration over $\Omega(x)$ is redundant, 
since by a change of variables
($U_\Omega(x,\mu):=\Omega(x)^\dagger U(x,\mu)\Omega(x+\mu))$ 
we can factor out the volume of the group. $Z_L[\beta, m^2,N]$ becomes
\begin{eqnarray}
\!\!\!\!\!\!
 \Big[\sum_{\{\Omega\}}\Big]\sum_{\{U\}} \exp -\frac{\beta}{2} 
\Big({\mathfrak Re}\sum_\Box Tr\big\{1- U_\Box\big\}
+ m^2 {\mathfrak Re}\sum_{x\mu}Tr\Bigl\{1- U(x,\mu)\Bigr\}\Big).
\label{sim.2}
\end{eqnarray}
However, in eq.(\ref{sim.1}) we force the integration over the
gauge orbit $U_\Omega$  by means of the explicit sum over  $\Omega$.
In doing this, we gain an interesting theoretical setup of the model;
in practice, our formalism is fully gauge invariant.
Moreover, by forcing the integration over the gauge orbit $U_\Omega$
we get  results which are  less noisy than those obtained by using only
the integration over the link variables in eq.(\ref{sim.2}). 
%
%

%%%%%%%%%%%%%%%%%%%%%%%%%%%%%%%%%%%%%%%%%%%%%%
\subsection{Numerical results}
\begin{figure}
\epsfxsize=100mm
\centerline{\epsffile{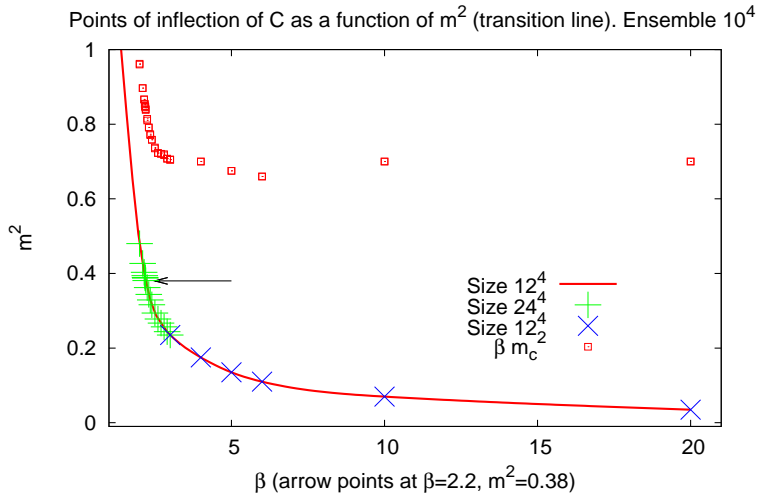}}
\caption{The transition line. The arrow marks the position of the end
point. {In the figures of the present paper the statistical errors are not displayed
since they are too small to be shown.}}
\label{fig.1}
\end{figure}

%%%%%%%%%%%%%%%%%%%%%%%%%%%%%%%%%%%%%%%%%%%%%%
%
The numerical analysis of the model \cite{Ferrari:2011aa} shows 
the existence
of a line (Fig.\ref{fig.1}) where the functions $\cal E$ and $\mathfrak C$ have 
inflection points in $m^2$ and $\beta$. In the region $\beta  >2.2$
the line separates the unconfined phase with vector
mesons from the phase where confinement may occur. 
%%%%%%%%%%%%%%%%%%%%%%%%%%%%%%%%%%%%%%%%%%%%%%%%%%%%%%
\par
\begin{figure}[p]
\begin{center}
\includegraphics[clip,width=1\textwidth,,height=0.4\textheight]
{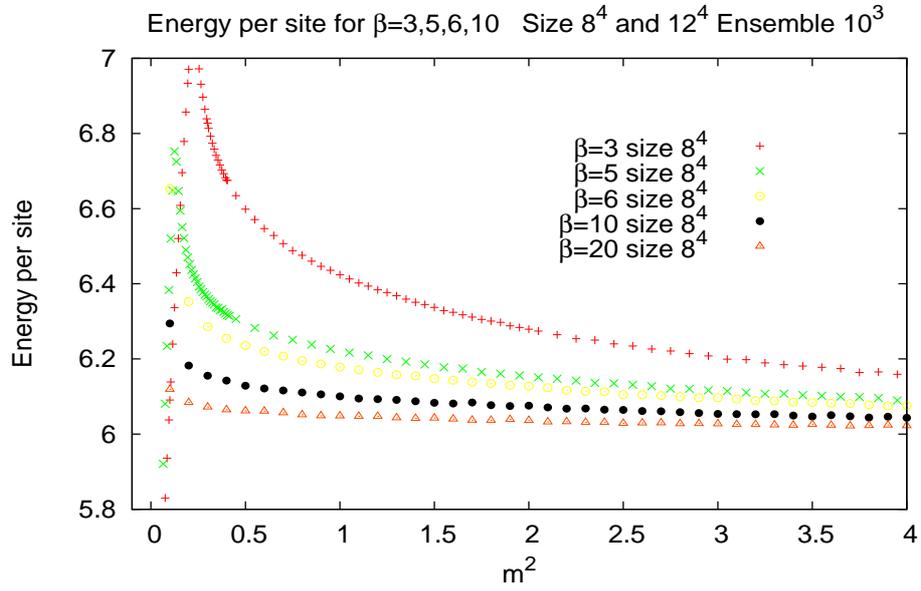}
\end{center}
\caption{Energy per site at $\beta=3,5,6,10,20$ as function
of $m^2$.}
\label{fig.2}
\end{figure}
\begin{figure}[p]
\begin{center}
\includegraphics[clip,width=1\textwidth,,height=0.4\textheight]
{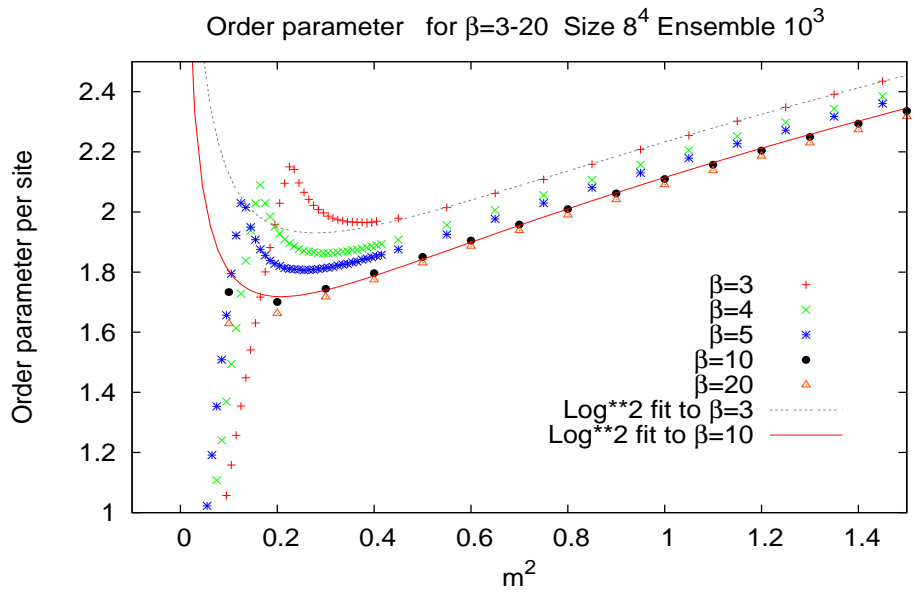}
\end{center}
\caption{Order parameter ${\mathfrak C}$ per site at $\beta=3,4,5,10,20$ as function
of $m^2$.}
\label{fig.2p}
\end{figure}

In the region of weak coupling (large $\beta$) and above the 
transition line we have evaluated the energy of the vacuum  $\cal E$ and
the order parameter
$\mathfrak C$ per site for
$m^2 =0.1 \dots 8$. We have used a cubic lattice with size
$8^4$ and $12^4$. No difference could be spotted in the results
for the two choices. We have taken $10^3$ measures separated by
$15$ updatings. For $\beta=10$ we have performed $10^4$ measures
in order to reduce the statistical errors, but no appreciable
improvement was  observed. Some results are given in Figs.\ref{fig.2}
and \ref{fig.2p}.
It is very interesting that both these global quantities 
converge for large $m^2$ 
to  values independent from $\beta$. From eq. (\ref{lat.2})
one would expect a strong dependence from the mass.
From both figures, we see that near the transition line the
dependence on $\beta$ becomes strong. In Fig. \ref{fig.2p} we
show a simple fit of the data at $\beta=3,10$ with
polynomials of second degree in $\ln(m^2)$.
%

%%%%%%%%%%%%%%%%%%%%%%%%%%%%%%%%%%%%%%%%%%%%%%%%%%
%
%

%%%%%%%%%%%%%%%%%%%%%%%%%%%%%%%%%%%%
%%%%%%%%%%%%%%%%%%%%%%%%%%%%%%%%%%%%
\section{The Fit}
\label{sec:fit}
%%%%%%%%%%%%%%%%%%%%%%%%%%%%%%%%%%%
%
In the present Section we try a fit of  the analytic results
of Section \ref{sec:YM} to the data obtained by simulation in
Section \ref{sec:sim}. We use a simplified form of the
mapping (\ref{intro.2})
\begin{eqnarray}&&
g^2 = f(\beta)
\nonumber\\&&
\frac{M^2}{\Lambda^2} ={m^2} \, {s(\beta)}
\nonumber\\&&
aM= t(\beta),
\label{intro.2p}
\end{eqnarray}
where the numerical values of $f,s,t$ are determined by the fit. 
The results of  Section \ref{sec:YM} become
\begin{eqnarray}&&
{\cal E}_M^{(2)} = \frac{3}{2} g^2 \frac{a^4M^4}{(4\pi)^4} 8.625 \Big[
 \Big(\ln m^2 + \ln s  - 4.939\Big)^2  -5.943 \Big] + e_0
\nonumber\\&&
{\mathfrak C}_M^{(0)}=
\frac{3}{2} \frac{a^4M^4}{(4\pi)^{2}} 8.625  
\Big[0.348 \Big(\ln m^2 + \ln s  -2.954 \Big) + 0.232 \Big] 
\nonumber\\&&
{\mathfrak C}_M^{(2)}=
\frac{3}{2}g^2\frac{a^4M^4}{(4\pi)^4} 8.625 \Big[
-2\Big(\ln m^2 +\ln s - 4.439 \Big)^2 + 12.386 \Big].
\label{fit.1}
\end{eqnarray}
The parameters are fitted according {to} the following steps:
i) first we fit the energy, since it is ${\cal O}(g^2)$;
ii) then we enter the parameters obtained by i) ({$[g^2 (aM)^4]$} 
and $\ln s$)
into ${\mathfrak C}_M$ and we perform the best fit for
\begin{eqnarray}
{\mathfrak C}_M^{(0)} + {\mathfrak C}_M^{(2)} + e_1 
\label{fit.2}
\end{eqnarray}
on a range of $m^2$ far from the transition line. The output of
the fit procedure are the parameters $g^2,aM, \ln s, e_0, e_1$.
For convenience, we shall display the quantity
\begin{eqnarray}
\beta_{\rm fit} \equiv \frac{4}{g^2},
\label{fit.2p}
\end{eqnarray}
where $g^2$ is the value obtained from the fit, 
in order to have a prompt comparison with the lattice parameter
$\beta$.
\par\noindent
The strategy is dictated by the fact that the direct fit of the order parameter
does not determine properly the values of $aM$ and $\ln s$,
since these last parameters have weak effect in comparison with the ${\cal O}(g^0)$
term. Thus it is better to work at first with
a ${\cal O}(g^2)$ quantity: the vacuum energy. However, the results of the first step depend
strongly on the range of $m^2$; therefore, we choose the interval
that yields a common value for $\ln s$ according to the parameterization
of eq. (\ref{intro.2p}). The interval is reported in Table
\ref{tab:energy} in the second column.
% 
%%%%%%%%%%%%%%%%%%%%%%%%%%%%%%%%%%%%%%%%%
\subsection{Fit of the Vacuum Energy}
\begin{figure}[p]
\begin{center}
\includegraphics[clip,width=1\textwidth,height=0.4\textheight]
{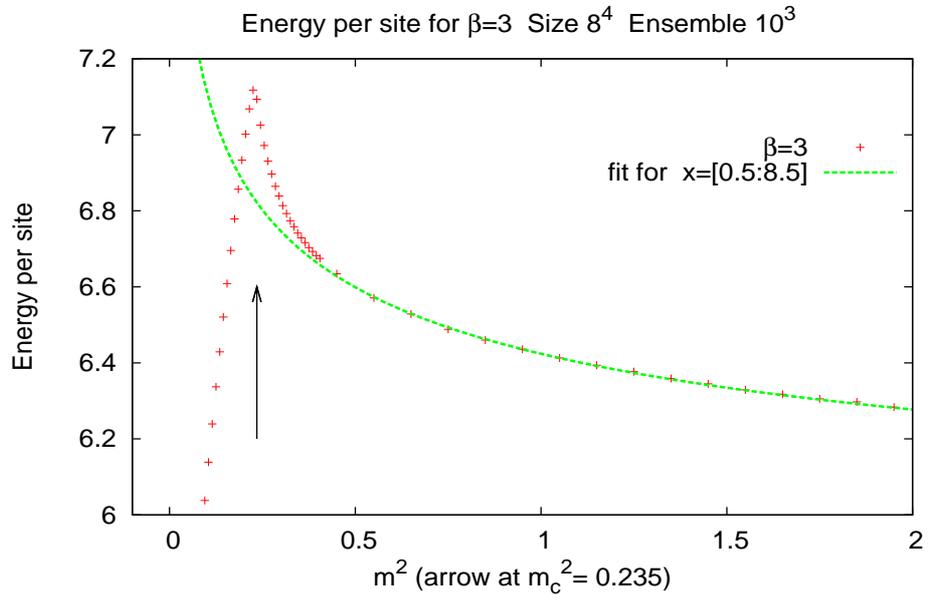}
\vspace*{-0.7 cm}
\end{center}
\caption{Fit  of the field theory prediction (eq. (\ref{fit.1})) to the lattice data for
$\beta=3$. The parameters are $g^2(aM)^4$, $\ln s$, $e_0$.}
\label{fig.3.1}
\end{figure}

\begin{figure}[p]
\begin{center}
\includegraphics[clip,width=1\textwidth,,height=0.4\textheight]
{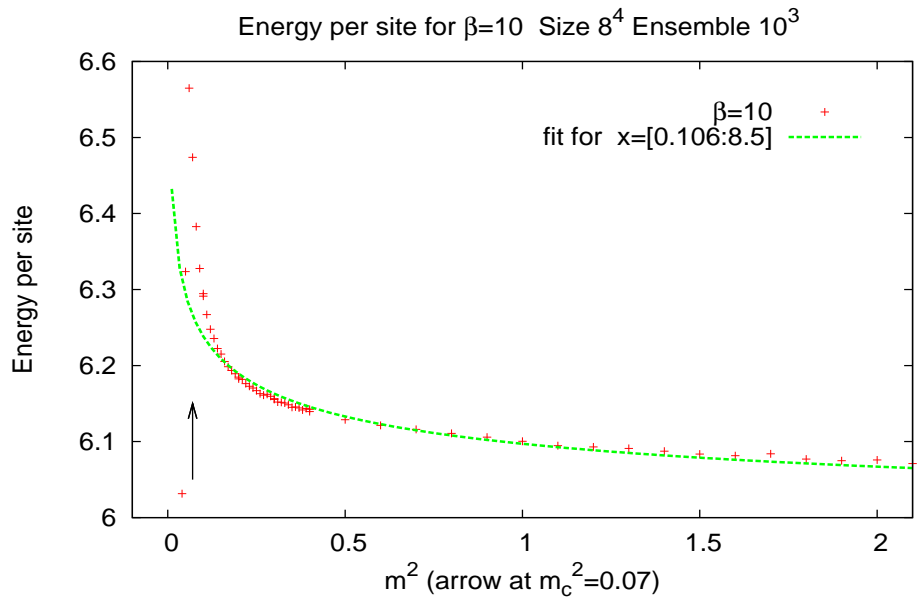}
\vspace*{-0.7 cm}
\end{center}
\caption{{See caption of Fig. \ref{fig.3.1}} for $\beta=10$.}
\label{fig.5}
\end{figure}

%%%%%%%%%%%%%%%%%%%%%%%%%%%%%%%%%%%%%%%%%
%
We perform the fit of the total energy ${\cal E}_M^{(2)}$ 
of eq. (\ref{fit.1}) for $\beta=3,4,5,6,7,8,9,$ $10,20$. 
Figs. \ref{fig.3.1} and \ref{fig.5} give a reasonable account 
of the match between lattice
gauge theory and the field theory calculation with our subtraction
procedure. 
In Table \ref{tab:energy} we list some of the data of the fit:
in particular $\ln s$, the $m^2$ interval and  $\chi^2$ (which is very small).
 It should be noticed that $aM$ and $g^2$ can be determined
only by using the fit of the order parameter, since the energy expression
contains only the product $[g^2 (aM)^4]$. 
%
%%%%%%%%%%%%%%%%%%%%%%%%%%%%%%%%%%%%%%%%%%%%%%%%%%%%%%%%%%%%%%%%%%%%%%
\subsection{Combined Fit of the Vacuum Energy and the Order Parameter}
\begin{figure}[p]
\begin{center}
\includegraphics[width = 0.9\textwidth]{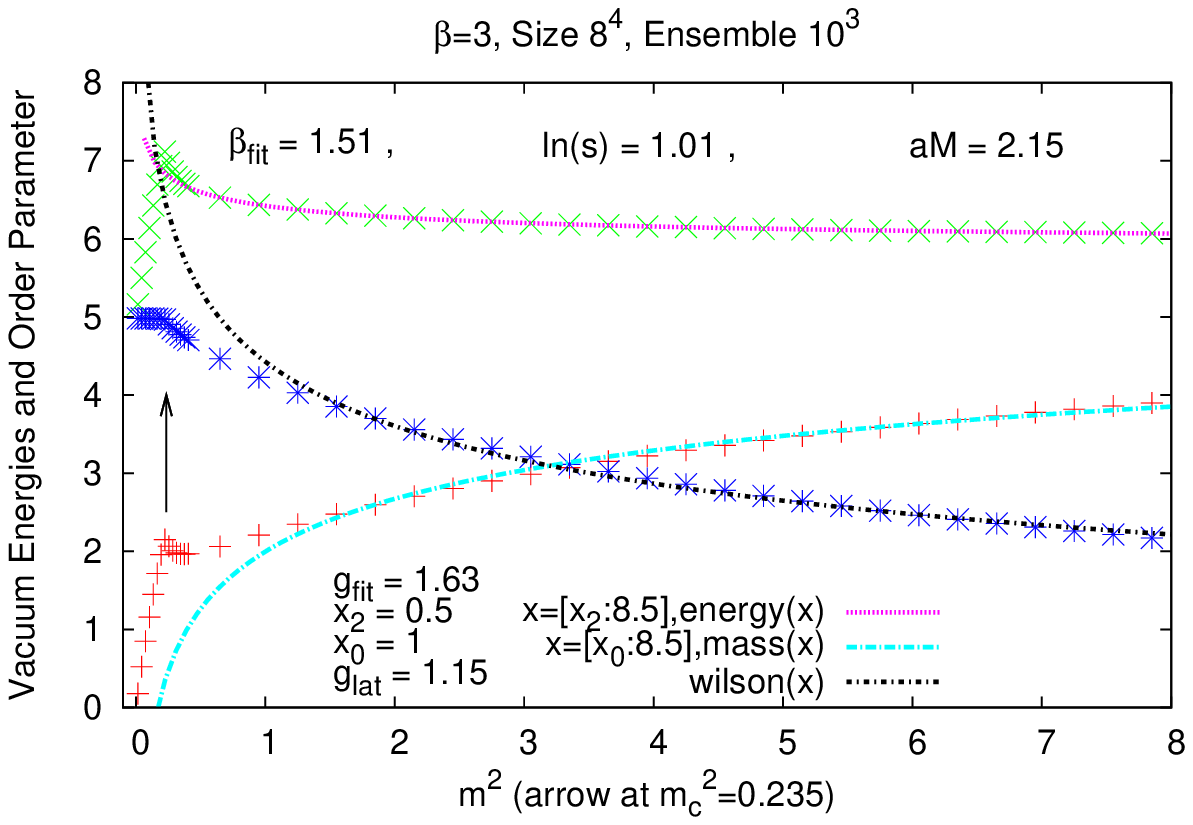}
\vspace*{-0.7 cm}
\end{center}
\caption{Fit of the field theory predictions (eqs. (\ref{fit.1}), (\ref{fit.2})) to the lattice data 
on vacuum energy and order parameter for $\beta=3$. { The variables in the order parameter 
are $aM$ and $e_1$, while $g^2(aM)^4$ and $\ln s$  are imported from the fit of the vacuum energy.}}
\label{fig.6}
\end{figure}

\begin{figure}[p]
\begin{center}
\includegraphics[width = 0.9\textwidth]{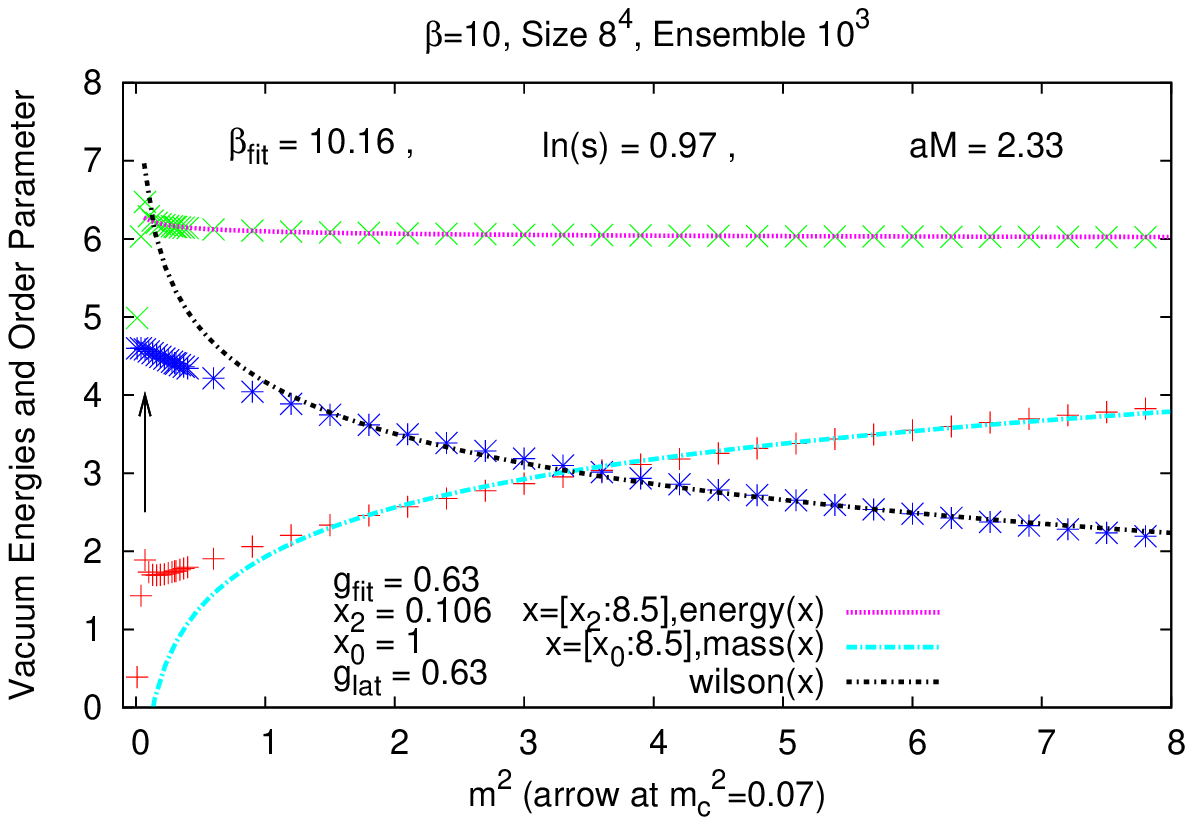}
\vspace*{-0.7 cm}
\end{center}
\caption{{See caption of Fig. \ref{fig.6}} for $\beta=10$.}
\label{fig.7}
\end{figure}

%%%%%%%%%%%%%%%%%%%%%%%%%%%%%%%%%%%%%%%%%%%%%%%%%%%%%%%%%%%%%%%%%%%%
%
In the second step we fit the order parameter; thus we find
the value of $g^2$ and $aM$. The results of the  procedure are listed
in Table \ref{tab:order} ($aM$, $\beta_{fit}=4/g^2$  and $\chi^2$ of the fit). 
In Figs. (\ref{fig.6}) and (\ref{fig.7}) we depict the results
of the fit procedure for the values $\beta=3,10$.
The plots include also the value of the Wilson action
of the vacuum {(${\cal E}- {\mathfrak C}$)}. The figures
show explicitly that the fit procedure intentionally
excludes the region near the transition line (at $m_c^2$).
\par
The numerical results show that the field theory
two-loop calculation of the energy of the vacuum
and the order parameter provides a good fit of
the lattice data on the surface given by the parametric
equations (\ref{intro.2p}). The data in Tables \ref{tab:energy},
\ref{tab:order} indicate that $aM$ is almost constant 
for high values of $\beta$. Instead $\beta_{fit}$ shows a steady 
increase.
It should be noticed that for high values of $\beta$ the
errors on the fit parameters become larger and larger.
\par
The results of the fit are rather surprising. The {\sl fundamental
length} $a$ is provided by the mass of the continuous theory
\begin{eqnarray}
a \simeq 2.3  \, M^{-1},
\label{lat.31}
\end{eqnarray}
while the values of $m^2$ correspond to different
values of the scale $\Lambda$ of the subtraction
procedure of the divergences
\begin{eqnarray}
m^2 \simeq \frac{1}{3}
\frac{M^2}{\Lambda^2}.
\label{lat.31p}
\end{eqnarray}
Thus the fit  departs substantially from the na\"{\i}ve 
identification $m=Ma$ where $a$ is the lattice spacing. 
This seems to be the unavoidable
price for a mapping of the continuum theory equipped with
more scales than the lattice can accommodate. 
\vspace*{-0.7 cm}
\begin{table*}[p]
\caption{\label{tab:energy} The fit of energy of the vacuum 
according to (\ref{fit.1}). $\ln s$ is defined in eq. (\ref{intro.2p}).} 
\begin{center} 
\begin{tabular}{|c||c|c|c|}
\hline 
$\beta$ & fit range & ln s & $\chi^2$  \cr\hline
\hline
3 & [0.5:8.5] & 1.009 $\pm$ 0.047  & 4.3e-06 \cr
\hline
4 & [0.218:8.5] & 0.99 $\pm$ 0.16 & 7.8e-05 \cr
\hline
5 & [0.165:8.5] & 1.07 $\pm$ 0.23  & 1.1e-04 \cr
\hline
6 & [0.147:8.5] & 1.13 $\pm$ 0.26 & 9.3e-05 \cr
\hline
7 &  [0.125:8.5] & 1.28 $\pm$ 0.28 & 9.8e-05 \cr
\hline
8 &  [0.115:8.5] & 1.17 $\pm$ 0.29 & 7.2 e-05 \cr
\hline
9 &  [0.109:8.5] & 1.28 $\pm$ 0.30 & 6.7e-05  \cr
\hline
10 &  [0.106:8.5] & 0.97 $\pm$ 0.31  & 4.0e-05 \cr
\hline
20 &  [0.07:8.5] & 1.03 $\pm$ 0.25  & 1.6e-05\cr
\hline
\end{tabular}
\end{center}
\end{table*}
\begin{table*}[p]
\caption{\label{tab:order} The fit of the order parameter 
according to (\ref{fit.1}). $\beta_{fit}$ is defined in eq. (\ref{fit.2p}).} 
\begin{center} 
\begin{tabular}{|c||c|c|c|c|c|}
\hline 
$\beta$ & $ m_c^2$ & fit range & $a M$  & $\beta_{fit}$ &  $\chi^2$\cr
\hline\hline
3 & 0.235 & [1.0:8.5] & 2.151 $\pm$ 0.011  & 1.506 $\pm$ 0.038 & 0.003 \cr
\hline
4 & 0.175 & [1.0:8.5] & 2.2357 $\pm$ 0.0089  & 2.66 $\pm$ 0.13 & 0.003 \cr
\hline
5 & 0.135 & [1.0:8.5] & 2.2774 $\pm$ 0.0081 & 3.80 $\pm$ 0.25 & 2.7e-03 \cr
\hline
6 & 0.11 & [1.0:8.5] & 2.2906 $\pm$ 0.0082 & 4.87 $\pm$ 0.35 & 2.8e-03 \cr
\hline
7 & 0.1 & [1.0:8.5] & 2.3067 $\pm$ 0.0079 & 5.82 $\pm$ 0.47 & 2.7e-03\cr
\hline
8 & 0.08 & [1.0:8.5] & 2.3155 $\pm$ 0.0078 & 7.17 $\pm$ 0.58  & 2.6e-03 \cr
\hline
9 & 0.07& [1.0:8.5] & 2.3237 $\pm$ 0.0077 & 8.13 $\pm$ 0.69 & 2.6e-03  \cr
\hline
10 & 0.06 & [1.0:8.5] & 2.3273 $\pm$ 0.0075 & 10.16 $\pm$ 0.83 &  2.5e-03  \cr
\hline
20 & 0.035 & [1.0:8.5] & 2.3492 $\pm$ 0.0051 & 21.27  $\pm$ 1.34 &  2.4e-03 \cr
\hline
\end{tabular}
\end{center}
\end{table*}
%

%%%%%%%%%%%%%%%%%%%%%%%%%%%%%%%%%%%%%%%%%%%%%%
%%%%%%%%%%%%%%%%%%%%%%%%%%%%%%%%%%%%%%%%%%%%%%
\vspace*{0.7 cm}
\section{Conclusions}
%%%%%%%%%%%%%%%%%%%%%%%%%%%%%%%%%%%%%%%%%%%%%%%
%
We have investigated the deconfined phase of 
a massive Yang-Mills model. In particular, we have
considered the weak coupling limit (large $\beta$ and $m^2 > m_c^2$)
of two global observables: the vacuum
energy density and the order parameter density.
The lattice simulation data (dependent on $\beta$
and $m^2$) have been fitted with the analytic 
two-loop calculations of the continuum field theory
(dependent on the mass $M$, the ultraviolet cut-off
$\Lambda$ and the lattice spacing $a$). The fit
turns out to be very promising both in supporting the 
conjecture of the lattice as a regulator for 
nonlinear gauge theories and in the comprehension
of the lattice parameters in phenomenology.
\par
After these results a tantalizing question remains: whether a
higher loop calculation (3 loop) of the energy and order
parameter density would improve the
agreement between lattice and continuum perturbation theory.
\par
The outlook is for a comparison of the lattice
data on the energy gaps with the radiative correction
of the masses (self-energies). Moreover, the
lattice simulations might provide quantitative predictions
near the phase transition line \cite{Ferrari:2011bx}.

%%%%%%%%%%%%%%%%%%%%%%%%%%%%%%%%%%%%%%%%%%%%%%
%%%%%%%%%%%%%%%%%%%%%%%%%%%%%%%%%%%%%%%%%%%%%%
\section*{Acknowledgements}
%%%%%%%%%%%%%%%%%%%%%%%%%%%%%%%%%%%%%%%%%%%%%
%
One of us (RF) is honored to thank the warm hospitality of the
Center for Theoretical Physics at MIT, Massachusetts, where part
of the present work  has been done.

\par\noindent

\begin{thebibliography}{99}

%\cite{Ferrari:2011aa}
\bibitem{Ferrari:2011aa}
  R.~Ferrari,
 % ``On the Phase Diagram of Massive Yang-Mills,''
  Acta Phys.\ Polon.\ B {\bf 43} (2012) 1965
  [arXiv:1112.2982 [hep-lat]].
  %%CITATION = ARXIV:1112.2982;%%

  
%\cite{Baaquie:1992cw}
\bibitem{Baaquie:1992cw} 
  B.~E.~Baaquie and G.~Bhanot,
  %``Microcanonical simulation of the four-dimensional SU(2) nonlinear sigma model,''
  Nucl.\ Phys.\ B {\bf 382}, 409 (1992).
  %%CITATION = NUPHA,B382,409;%%


%\cite{Fradkin:1978dv}
\bibitem{Fradkin:1978dv}
  E.~H.~Fradkin and S.~H.~Shenker,
  %``Phase Diagrams Of Lattice Gauge Theories With Higgs Fields,''
  Phys.\ Rev.\  D {\bf 19}, 3682 (1979).
  %%CITATION = PHRVA,D19,3682;%%


%\cite{Jersak:1985nf}
\bibitem{Jersak:1985nf}
  J.~Jersak, C.~B.~Lang, T.~Neuhaus, G.~Vones,
  %``Properties Of Phase Transitions Of The Lattice SU(2) Higgs Model,''
  Phys.\ Rev.\  {\bf D32}, 2761 (1985).
  

%\cite{Evertz:1985fc}
\bibitem{Evertz:1985fc}
  H.~G.~Evertz, J.~Jersak, C.~B.~Lang, T.~Neuhaus,
  %``Su(2) Higgs Boson And Vector Boson Masses On The Lattice,''
  Phys.\ Lett.\  {\bf B171}, 271 (1986).
  

%\cite{Evertz:1986vp}
\bibitem{Evertz:1986vp}
  H.~G.~Evertz, V.~Grosch, J.~Jersak, H.~A.~Kastrup, T.~Neuhaus, D.~P.~Landau, J.~L.~Xu,
  %``Monte Carlo Analysis Of Gauge Invariant 2 Point Functions In An Su(2) Higgs Model,''
  Phys.\ Lett.\  {\bf B175}, 335 (1986).
  


%\cite{Campos:1997dc}
\bibitem{Campos:1997dc}
  I.~Campos,
  %``On the SU(2)-Higgs phase transition,''
  Nucl.\ Phys.\  B {\bf 514}, 336 (1998)
  [arXiv:hep-lat/9706020].
  %%CITATION = NUPHA,B514,336;%%


%\cite{Greensite:2006ng}
\bibitem{Greensite:2006ng}
  J.~Greensite and S.~Olejnik,
  %``Vortices, symmetry breaking, and temporary confinement in SU(2)
  %gauge-Higgs theory,''
  Phys.\ Rev.\  D {\bf 74}, 014502 (2006)
  [arXiv:hep-lat/0603024].
  %%CITATION = PHRVA,D74,014502;%%

%\cite{Caudy:2007sf}
\bibitem{Caudy:2007sf}
  W.~Caudy and J.~Greensite,
  %``On the Ambiguity of Spontaneously Broken Gauge Symmetry,''
  Phys.\ Rev.\  D {\bf 78}, 025018 (2008)
  [arXiv:0712.0999 [hep-lat]].
  %%CITATION = PHRVA,D78,025018;%%



%\cite{Bonati:2009pf}
\bibitem{Bonati:2009pf}
  C.~Bonati, G.~Cossu, M.~D'Elia and A.~Di Giacomo,
  %``Phase diagram of the lattice SU(2) Higgs model,''
  Nucl.\ Phys.\  B {\bf 828}, 390 (2010)
  [arXiv:0911.1721 [hep-lat]].
  %%CITATION = NUPHA,B828,390;%%


%\cite{Bettinelli:2007tq}
\bibitem{Bettinelli:2007tq}
  D.~Bettinelli, R.~Ferrari and A.~Quadri,
  %``A Massive Yang-Mills Theory based on the Nonlinearly Realized Gauge
  %Group,''
  Phys.\ Rev.\  D {\bf 77} (2008) 045021
  [arXiv:0705.2339 [hep-th]].

%\cite{Bettinelli:2007cy}
\bibitem{Bettinelli:2007cy}
  D.~Bettinelli, R.~Ferrari and A.~Quadri,
  %``One-loop Self-energy and Counterterms in a Massive Yang-Mills Theory based
  %on the Nonlinearly Realized Gauge Group,''
  Phys.\ Rev.\  D {\bf 77}, 105012 (2008)
  [arXiv:0709.0644 [hep-th]].
  %%CITATION = PHRVA,D77,105012;%%
  
%\cite{Ferrari:2005ii}
\bibitem{Ferrari:2005ii}
  R.~Ferrari,
  %``Endowing the nonlinear sigma model with a flat connection structure: A  way
  %to renormalization,''
  JHEP {\bf 0508}, 048 (2005)
  [arXiv:hep-th/0504023].
  %%CITATION = JHEPA,0508,048;%%


%\cite{Ferrari:2005va}
\bibitem{Ferrari:2005va}
  R.~Ferrari and A.~Quadri,
  %``A weak power-counting theorem for the renormalization of the non-linear
  %sigma model in four dimensions,''
  Int.\ J.\ Theor.\ Phys.\  {\bf 45}, 2497 (2006)
  [arXiv:hep-th/0506220].
  %%CITATION = IJTPB,45,2497;%%



%\cite{Bettinelli:2007zn}
\bibitem{Bettinelli:2007zn}
  D.~Bettinelli, R.~Ferrari and A.~Quadri,
  %``Further comments on the renormalization of the nonlinear sigma model,''
  Int.\ J.\ Mod.\ Phys.\  A {\bf 23}, 211 (2008)
  [arXiv:hep-th/0701197].
  %%CITATION = IMPAE,A23,211;%%
 

%\cite{Streater:1989vi}
\bibitem{Streater:1989vi} 
  R.~F.~Streater and A.~S.~Wightman,
  %``PCT, spin and statistics, and all that,''
  Princeton, USA: Princeton Univ. Pr. (2000) 207 p.

%\cite{Davydychev:1995mq}
\bibitem{Davydychev:1995mq} 
  A.~I.~Davydychev and J.~B.~Tausk,
  %``A Magic connection between massive and massless diagrams,''
  Phys.\ Rev.\ D {\bf 53}, 7381 (1996)
  [hep-ph/9504431]. 
%  {\bf   }
  %%CITATION = HEP-PH/9504431;%%


%\cite{Wilson:1974sk}
\bibitem{Wilson:1974sk}
  K.~G.~Wilson,
  %``CONFINEMENT OF QUARKS,''
  Phys.\ Rev.\  D {\bf 10}, 2445 (1974).
  %%CITATION = PHRVA,D10,2445;%%
  

%\cite{Ferrari:2011bx}
\bibitem{Ferrari:2011bx}
  R.~Ferrari,
  %``Metamorphosis versus Decoupling in Nonabelian Gauge Theories at Very 
  %High Energies,''
  Acta Phys.\ Polon.\ B {\bf 43} (2012) 1735
  [arXiv:1106.5537 [hep-ph]].
  %%CITATION = ARXIV:1106.5537;%%

\end{thebibliography}
\end{document}